# Neural, Muscular, and Perceptual responses with Shoulder Exoskeleton Use over Days

Rana Mukherjee, Tiash[1], Tyagi, Oshin[2], Wang, Jingkun[3], Kang, John J.[2], Mehta, Ranjana K.[1,2]

[1]Department of Mechanical Engineering, Texas A&M University
[2]Department of Industrial and Systems Engineering, Texas A&M University
[3]Department of Industrial and Systems Engineering, Purdue University

**SYNOPSIS**

Passive shoulder exoskeletons have been widely introduced in the industry to aid upper extremity movements during repetitive overhead work. As an ergonomic intervention it is important to understand how users adapt to these devices over time and if these induce external stress while working. The study evaluated the use of exoskeleton over a period of 3 days, by assessing the neural, physiological and perceptual responses of twenty-four participants by comparing a physical task against the same task with additional cognitive workload. Over days adaptation to task irrespective of task and group were identified. Electromyography (EMG) analysis of shoulder and back muscles reveal lower muscle activity in exoskeleton group irrespective of task. Functional Connectivity analysis using functional near infrared spectroscopy (fNIRS) reveals exoskeletons benefit the users by reducing task demands in the motor planning and execution regions. Sex based differences were also identified in these neuromuscular assessments.

**BACKGROUND**

A majority of the cases of Musculoskeletal Disorders (MSDs) in the United States involve workers employed in manufacturing, and social assistance workforce. Repetitive overhead work, that is work done over the acromion level, is one of the leading causes of work-related MSDs [1]. This causes increased absenteeism but also affects the worker's wellbeing and health [2, 3]. Despite automation, most tasks require overhead work with tools, installing, and repairing components [4]. The risks associated with supporting the arm weight increases physical stress on the upper body muscles due to weight for longer periods, in awkward positions increasing the risk of muscle fatigue due to insufficient rest and recovery [5].

Industrial occupational exoskeletons are wearable devices, worn over the body, and they assist and augment a worker's movements. Passive shoulder exoskeletons support the weights of the arms, reducing strains on the shoulder muscles by distributing the load onto the back muscles. These exoskeletons rely on harvesting the potential energy stored in the springs of the exoskeletons due to body weight[6].

Studies evaluating exoskeletons discuss biomechanical benefits but, exoskeletons also introduce new risks. Motion adaptation, unwanted discomfort due to limited range of motion, and higher cognitive load may lower workers' perceptions of exoskeleton utility and result in unintended consequences [7]. This study aimed to examine how exoskeleton use and its potential benefits vary over time, which would further aid in developing better metrics to improve the use and implementation of exoskeletons. A second aim was to evaluate how the worker performs the same tasks with the introduction of cognitive workload. It was hypothesized that improvements in neural, perceptual, and muscular responses will be noted as the worker adapts to the exoskeleton, but differences will be noticed with the introduction of cognitive tasks.

**METHODOLOGY**

Gender balanced 24 participants were randomly assigned into 2 groups: Control and Exoskeleton groups. The two tasks included: Single task – participants only performed the overhead reaching and pointing task; and Dual task – participants were asked to perform the single task while simultaneously subtracting 13 from a four-digit number randomly given to them during the experiment. Each trial consisted of 24 button presses. The tasks were counterbalanced during the experiment. The study protocol (Figure 1(a)) included training on Day 1 to familiarize the participants with the task.

The between subject independent variables were sex (male, female), group (control, exoskeleton). The within subject independent variables were day (day1, day2, day3), task (single, dual). Participants in exoskeleton group performed the experiment wearing Eksovest (Ekso Bionics Holdings Inc., CA, USA), which was adjusted for each participant for their body segment dimension.

Task performance was evaluated based on time taken to complete each trial. Physical and mental workload were measured using Borg's Rate of Perceived Exertion ( RPE) scale and NASA TLX, respectively [8, 9].

Electromyography (EMG) was collected from the participants muscles (R/L biceps – R/LBi, R/L medial triceps – R/LMT , R/L lateral triceps – R/LLT, R/L anterior deltoid – R/LAD , R/L middle deltoid – R/LMD, R/L upper trapezius – R/LUT, and R/L lumbar erector spinae – R/LES. Functional near infrared spectroscopy (fNIRS) custom probe design was employed to obtain temporal correlation between functionally independent regions in the brain, namely,  L/R prefrontal cortex (L/RPFC), supplementary motor area (SMA), L/R premotor cortex (L/RPM), and L/R primary motor area (L/RM1). The probe design used for this analysis has been described in Figure 1(b). Heart Rate (HR), Root Mean Square of the Successive Differences (RMSSD) and the Low Frequency/High Frequency Ratio (LF/HF ratio) were taken. Repeated measures analysis of variance was performed for statistical analysis.

**RESULTS**

**Task Performance and Subjective Response**

Day effect [$p < 0.001$] for performance showed improvement over days. Single tasks showed better performance than dual tasks [$p < 0.001$]. Day×Task [$p < 0.001$] showed time taken to complete the single task was lower than time taken to complete the dual task on all days. Day×Task×Sex [$p = 0.019$] showed significant results for females.  On day 1 time taken to complete single task was lower than time taken for dual task. Times for dual task were significantly higher for females on day 1 than the other days. However, for males, time to completion for dual task was significantly lower only on Day 3 as compared to on day 1. A four-way interaction effect of Day×Task×Group×Sex [$p < 0.009$] showed that for the Exoskeleton group, on day 1, females significantly took more time to complete the dual task than they took for the single task. For the dual task, females took longer to complete the task on day 1 as compared to the other days. For the dual tasks in the control group, both sexes on day 3 took significantly lower time to complete task than on day 1. A between subject main effect of Group was noted in RPE [$p = 0.046$], where Exoskeleton group reported lower RPE than control group. Day effect noted [$p = 0.002$], where TLX was greater for day 1 as compared to the other days. Task effect [$p < 0.001$] was also noted where scores were higher for Dual task. An interaction effect of Task×Sex [p = 0.049] was noted where scores for dual task were higher for both the sexes, but more in females.

**Heart rate and Heart rate variability**
Dual tasks resultes in higher HR than single tasks [p = 0.006]. Day×Group×Sex [p = 0.049] showed males in control group had higher HR than in exoskeleton group on day 1 but for day 3 both sexes had higher HR for exoskeleton group. Task effect [p = 0.046] seen in LFHF ratio showed dual task reported higher values for both groups. No significant effects were noted in RMSSD.

**Muscle Activity**

Day effect in RUT [p = 0.001] was observed, where day 2 had significantly higher activity as compared to the other days and group effect in RUT [p = 0.015], RMD [p = 0.008] and RAD [p = 0.003] where control group had higher muscle activity. Task effect seen in RLT [p = 0.030], RBi [p = 0.047], LES [p < 0.001], RES [p = 0.039], RUT [p = 0.001], LMD [p = 0.019], RMD [p = 0.023] and RAD [p = 0.025] showed higher activity during single task. Sex effect in LLT [p = 0.019], LES [p < 0.001] and RUT [p = 0.015] showed females exhibited higher muscle activity than the males. However, Group×Sex in RUT [p = 0.005] showed females in control group had greater activity than those in exoskeleton group. Males however reported lower muscle activity than females in the control group.

**Neural Activity**

Group effect on the Functional Connectivity (FC) between the following regions was identified L/RPFC [p = 0.004], LPFC-RPM [p = 0.004], LPFC-LM1 [p = 0.029], RPFC-RM1 [p = 0.043], LPM-LM1 [p = 0.042], RPM-LM1 [p = 0.01] and RPM-RM1 [p = 0.015] where control group exhibited stronger connection. Sex effect in L/R PFC [p = 0.008] and SMA-RPM [p = 0.021] showed males exhibited stronger FC. Task effect was noted in RPFC-RM1 [p = 0.028], where Single task exhibited stronger connection but in LPM-RPM [p = 0.035] Dual task exhibited stronger FC. Day×Task in SMA-LM1 [p = 0.003] showed on day 3, FC in single task was significantly greater than dual task but for RPM-LM1 [p = 0.019], where day 2 were lower Group×Sex in LPFC-RPM [p = 0.002] revealed FC for Males in Exoskeleton group was weaker and sex differences were noticed in control group where Females had weaker connection. Group×Task interaction effect in LPM-RPM [p = 0.01] revealed for control, single task had weaker FC. Sex×Task interaction in LPFC-RM1 [ p = 0.04] showed females had weaker FC than males in during Dual task Day×Sex×Task in LPFC-SMA [p = 0.045] showed for females, Single task had stronger FC than during dual task, however no such significant difference was found for the Males. Day×Sex×Task in LPFC-LM1 [p = 0.007] showed that FC went down on Day 3 as compared to Day 2 for males in control group. However, the same was not observed for Males in the exoskeleton group.

**DISCUSSION**

Adaptation to task over days both single and dual irrespective of the group, were seen in this study. However, performance was better for single task. Interestingly, statistically significant results were observed for females in exoskeleton group on Day 1 where more time was taken to complete the dual task as compared to the single task. Sex based differences were also observed in past studies conducted for Trunk exoskeletons, but only for physical tasks with exoskeletons, however our results show irrespective of group, females adapted better to the dual task [10]. Our results confirm that irrespective of tasks, perceived physical workload was lower for exoskeleton group. Interestingly, irrespective of group, main effects of day, sex and task were noted in TLX scores [11]. As required, for dual task greater scores were reported. Females reported much higher scores as compared to males for dual task thus verifying, the lower performance on day 1.

As hypothesised, exoskeletons reduced muscle stress on the shoulder muscles. However, for dual task muscle activity were lower. Comparing these results with performance it could be concluded perhaps due to division of attention to the mental math, the time taken to complete the task was more [12].

Our study finds weaker connectivity for the exoskeleton group, where it can be postulated that with the introduction of exoskeletons movement planning is not as intense as for without any support, proving that exoskeleton benefits the participant while working. The significantly stronger connectivity between, LPM which is more involved in motor planning and skill development, and RPM, which is majorly involved in working memory during dual task in control group while no significant connectivity in the exoskeleton group is a strong indication, that maybe with the introduction of exoskeletons, the high task demands have been lowered [14]. Further exploration based on sex differences would help understand the adaptation strategies adopted by the different genders. Neural evaluation would also help identify tasks which befit from these industrial interventions hence develop better training modules for improved use.

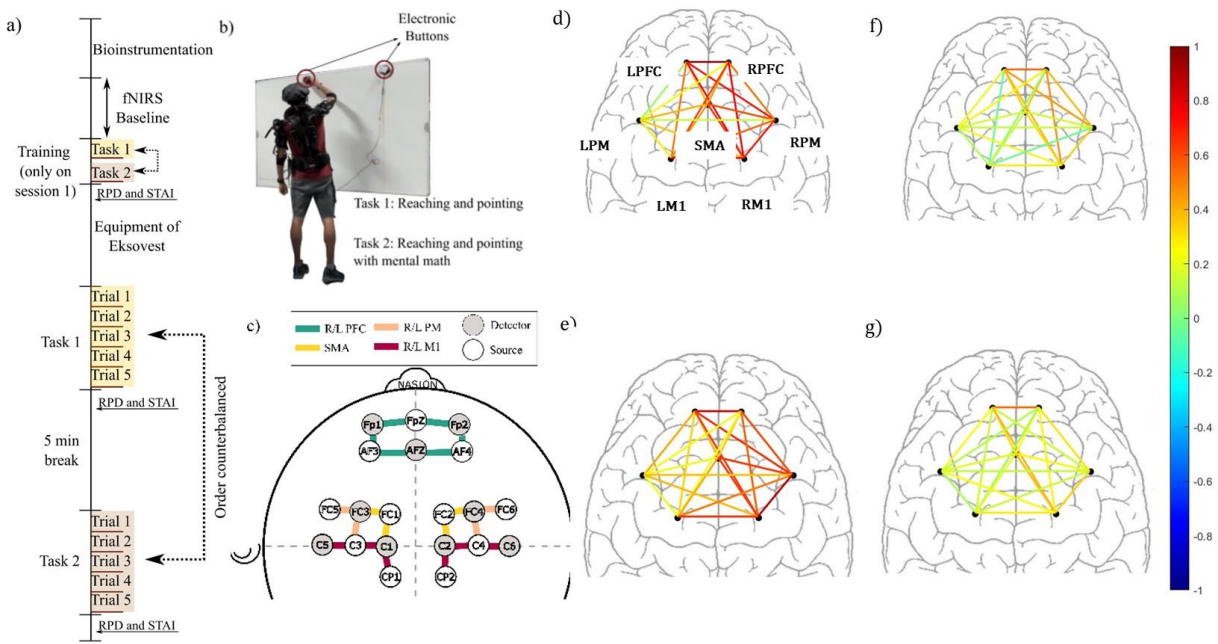

Figure 1: a) Study Protocol b) Experimental Set-up c) Probe-design for Neural Activity data collection using fNIRS; Functional Connectivity graphs - d) Control group while single task e) Control group while dual task f) Exoskeleton group while single task g) Exoskeleton group while dual task

**DECLARATION**

Nothing to disclose by any author(s).